\documentclass[%
 reprint,
 amsmath,amssymb,
 aps,
]{revtex4-2}

\usepackage{graphicx}% Include figure files
\usepackage{dcolumn}% eqnarray table columns on decimal point
\usepackage{bm}% bold math
\begin{document}

%\preprint{APS/123-QED}

\title{Approximation of human flow in urban areas by a network of electric circuits: Potential fields and fluctuation-dissipation relations}% Force line breaks with \\

\author{Yohei Shida$^{1}$}
\author{Jun'ichi Ozaki$^{2}$}%
\author{Hideki Takayasu$^{2,3}$}
\author{Misako Takayasu$^{1,2*}$}%

\affiliation{%
$^{1}$Department of Mathematical and Computing Science, School of Computing, Tokyo Institute of Technology, 4259 Nagatsuta-cho, Midori-ku, Yokohama 226-8503, Japan}%
\affiliation{%
$^{2}$Institute of Innovative Research, Tokyo Institute of Technology, 4259 Nagatsuta-cho, Midori-ku, Yokohama 226-8503, Japan}%
\affiliation{%
$^{3}$Sony Computer Science Laboratories, 3-14-13 Higashi-Gotanda, Shinagawa-ku, Tokyo, Japan}%

\date{\today}% It is always \today, today,

\maketitle

%%%%%%%OKKKKKKKKKKKKKKK
%\tableofcontents
%%\section{Introduction}
\textbf{Owing to the big data the extension of physical laws on nonmaterial has seen numerous successes, and human mobility is one of the scientific frontier topics. Recent GPS technology has made it possible to trace detailed trajectories of millions of people\cite{jeffrey2020anonymised,lutu2020characterization,pepe2020covid,gao2020association,yabe2020non,bonaccorsi2020economic,huang2020twitter,jia2020population}, macroscopic approaches such as the gravity law for human flow between cities\cite{ravenstein1885laws,noulas2012tale,thiemann2010structure,barthelemy2014spatial,simini2021deep} and microscopic approaches of individual origin-destination distributions\cite{gonzalez2008understanding,song2010limits,cuttone2018understanding,zhao2015explaining,jurdak2015understanding,alessandretti2017multi} are attracting much attention. However, we need a more general basic model with wide applicability to realize traffic forecasting and urban planning of metropolis fully utilizing the GPS data. Here, based on a novel idea of treating moving people as charged particles, we introduce a sophisticated method to map macroscopic human flows into currents on an imaginary electric circuit defined over a metropolitan area. Conductance is found to be nearly proportional to the maximum current in each location and synchronized human flows in the morning and evening are well described by the temporal changes of electric potential. Surprisingly, the famous fluctuation-dissipation theorem holds, namely, the variances of currents are proportional to the conductivities akin to an ordinary material. Especially during the pandemic, such a tool may offer an invaluable insight for policy-making in managing human flows.}

%%%%%%%%%%%%%kokoni
Services that use GPS data are now deeply rooted in our daily lives (e.g., car navigation systems and location-based smartphone games). A large amount of high-frequency data collected from GPS services are used for studies without infringing on people's privacy\cite{jeffrey2020anonymised,lutu2020characterization,pepe2020covid,gao2020association,yabe2020non,bonaccorsi2020economic,huang2020twitter,jia2020population}. Over the past decade, scientific research on human mobility using high-frequency location data can be categorized roughly into two categories: macroscopic analysis of human migration between cities typically based on gravity-like laws\cite{ravenstein1885laws,noulas2012tale,thiemann2010structure,barthelemy2014spatial,simini2021deep}, and microscopic modeling of individual movement patterns based on the origin-destination matrix\cite{gonzalez2008understanding,song2010limits,cuttone2018understanding,zhao2015explaining,jurdak2015understanding,alessandretti2017multi}. 

The third category of mesoscopic studies appeared recently, focusing on the collective flow of people in metropolitan areas. The human flux toward the city center during the morning rush hour was analyzed to estimate macroscopic driving force potentials\cite{mazzoli2019field}. Furthermore, drainage basin structures defined from the analogy of water flow of rivers are found to have scale-free properties\cite{shida2020universal,shida2021universal}. 

Herein, we analyze mesoscopic human mobility in urban areas based on a novel analogy of electric circuits, in which collective human movements are approximated by electric currents of an imaginary lattice network of registers defined over an entire urban area. The values of the estimated conductance reflect the infrastructure of transportation (e.g., railways), and the electric potentials are determined as the driving force of the mean human flux at each time interval. 

In the last part of this paper, we focus on the fluctuation around the mean flow caused by the electric potentials. For materials at thermal equilibrium, it is well known that the fluctuation width is determined by the strength of dissipation or resistivity, known as the fluctuation-dissipation relation\cite{nyquist1928thermal}. Recently, the applicability of this relation has been extended to nonmaterial fields (e.g., financial markets), based on the analogy to colloidal Brownian motion in fluid molecules\cite{yura2014financial,kanazawa2018derivation}. The motion of market price is described by the Langevin equation and the fluctuation-dissipation relation holds by analyzing high-quality market data of buy and sell orders, which correspond to molecules surrounding an imaginary colloidal particle. In this analogy, a colloidal particle corresponds to the market price. We will confirm the validity of the fluctuation-dissipation relation of our imaginary electric circuit model that describes the human flux in urban areas.

%%%%%%%OKKKKKKKKKKKKKKK
%%\section{Data, Method, and Result}
%%\subsection{Data}

The GPS data were provided by a Japanese company, Agoop\cite{Agoop53:online}, containing the location of smartphones of approximately a quarter of million users in Japan throughout 2015. The data (i.e., user ID, time, coordinates, and velocity) were recorded at approximately 30 min intervals from morning to midnight. To protect privacy, no data were recorded from midnight to morning, and the ID was randomized every morning. As the population of Japan is approximately 125 million, one user represents approximately 500 people.

%%%%%%MADA
%%\subsection{Current}
To make the analogy to electric circuits, we pay attention to moving people with nonzero velocity and regard them as positively charged particles; they are assumed to be driven by an electric field. Those people with zero velocity are treated as chargeless, thus the source and sink for the charge are established. For example, in the morning, residential and business areas act as sources and sinks, respectively. 

\begin{figure*}

\centering

\includegraphics{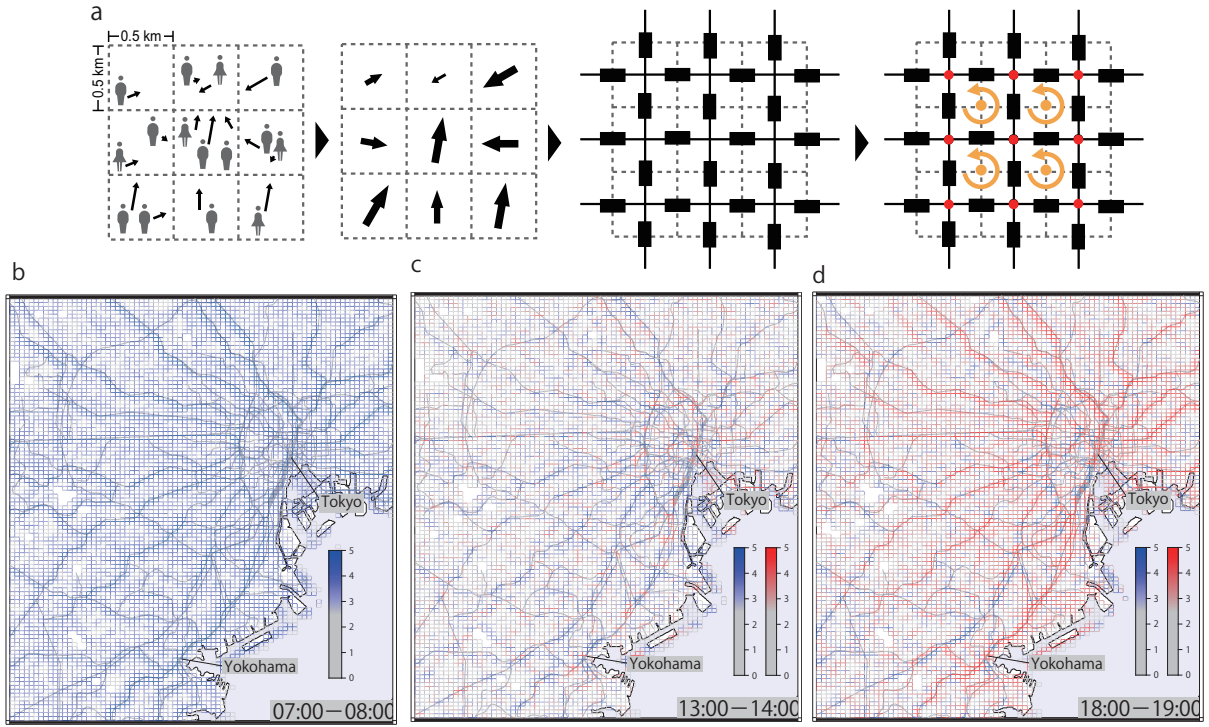}

\caption{\textbf{Definition of the imaginary electric circuit and observed currents.} (a) Schematic figures demonstrating the way we regard the human flow as electric current. The mean velocity of people in each cell is calculated by evaluating their average velocity components over those people who are moving in the square during the observation interval of 1 h. The current and resistance were defined between adjacent cells. Rotation is defined at the orange dots where the four cells meet, and the potential is defined at the red dots in the center of the cell. (b-d) Current patterns on the map in the morning, afternoon, and evening. The red lines show the currents with different direction compared to the morning current pattern. The strength of the flow was plotted on a logarithmic scale. The gray lines in the map represent the railways}
\label{fig:fig1}
\end{figure*}

We divide the map into a grid of 500 $\times$ 500 m squares and calculate the population density of the moving people $\overline{\rho}_{i,j,k}$ in the cell and their mean velocity $\overline{\bm{v}}_{i,j,k}$ every 1 hour from 5:00 to 24:00; $(i, j)$ denote the coordinates (longitude and latitude) on the map, and $k$ denotes the time interval of 1 h (see Figs. \ref{fig:fig1}a-left and \ref{fig:fig1}a-middle left)\cite{shida2020universal}. In this process, we averaged over all weekdays of 2015 to obtain enough data points in each cell for each hour. The longitudinal electric current  from cell $(i, j)$ to the neighbor cell $(i+1, j)$ at time interval $k$ is given by the following equation:
\begin{eqnarray}
\bm{I}_{(i,j)\rightarrow(i+1,j),k}=\frac{\overline{v}_{x,i,j,k}\overline{\rho}_{i,j,k}+\overline{v}_{x,i+1,j,k}\overline{\rho}_{i+1,j,k}}{2},
\label{eq:def-current}
\end{eqnarray}
where $\overline{v}_{x,i,j,k}$ is the longitudinal component of the velocity vector $\overline{\bm{v}}_{i,j,k}$; likewise for the latitudinal component. These currents are defined as dark rectangles in Fig. \ref{fig:fig1}a-middle right; resistance is defined later. The proposed electric circuit model is illustrated in Fig. \ref{fig:fig1}a-right. Red dot represents the center of a cell where the electric potential and the quantity of the sink and source of charges are defined. Yellow dots represent the corners of four cells where rotation of the electric field is introduced. 

Figs. \ref{fig:fig1}b ,\ref{fig:fig1}c and \ref{fig:fig1}d show examples of current strength maps around the center of Tokyo in the morning (7:00-8:00), afternoon (13:00-14:00), and evening (18:00-19:00). The color depth indicates the strength of current. Fig. \ref{fig:fig1}b shows a typical current pattern of morning rush hour where strong currents are observed along railroads and highways directing toward the city center. Fig. \ref{fig:fig1}c shows a typical afternoon current pattern where red indicates a current direction opposite to the morning pattern in Fig. \ref{fig:fig1}b. The overall strength of currents is weaker and about half of the directions differ from morning. The evening pattern Fig. \ref{fig:fig1}d are mostly in red, showing that the current directions are mostly opposite to the morning pattern.

The next step is to determine the values of resistance of our model based on the idea that the rotation of the electric field at each corner of the four cells, defined by the following equation, should be approximately zero.
\begin{eqnarray}
\nabla\times(\bm{I}R)_{i+0.5,j+0.5,k}=\,\,\,\,\,\,\,\,\,\,\,\,\,\,\,\,\,\,\,\,\,\,\,\, \nonumber \\
\frac{(\bm{I}R)_{(i+1,j)\rightarrow(i+1,j+1),k}-(\bm{I}R)_{(i,j)\rightarrow(i,j+1),k}}{\Delta x} \nonumber \\
-\frac{(\bm{I}R)_{(i,j+1)\rightarrow(i+1,j+1),k}-\bm{(I}R)_{(i,j)\rightarrow(i+1,j),k}}{\Delta y},
\label{eq:def-rot}
\end{eqnarray}
where the cell sizes in both directions are regarded as unity (i.e., $\Delta x=\Delta y=1$). We assume that the resistance values reflect the infrastructure of cities and do not depend on time. The values of resistance are determined iteratively using Adam\cite{kingma2014adam}---a type of steepest descent method to minimize the sum of the square of rotation for all locations at all times with the restriction that each resistance is positive while conserving the total sum of resistance, i.e.:
\begin{eqnarray}
&{\rm minimize} \,\,\,\, L=\sum_{i,j,k}   (\nabla\times(\bm{I}R^{\tau})_{i+0.5,j+0.5,k})^{2} \nonumber\\
&{\rm subject\,to} \,\,\,\, R_{(i,j),(i',j')}>0 \\
&\sum R_{(i,j),(i',j')}^{0}=\sum R_{(i,j),(i',j')}^{\tau}
\label{eq:optimization}
\,\,\,\,(R_{(i,j),(i',j')}^{0}=1), \nonumber
\end{eqnarray}
where $R^{\tau}$ denotes the resistance at the $\tau$-th iteration that starts with a uniform initial condition $R^0=1$. The number of resistors in this circuit model for urban areas of Tokyo is approximately 75000, and the resistance values of the resistors on the boundary of the model were fixed as infinity so that no flux enters through the boundary.

\begin{figure*}

\centering
\includegraphics{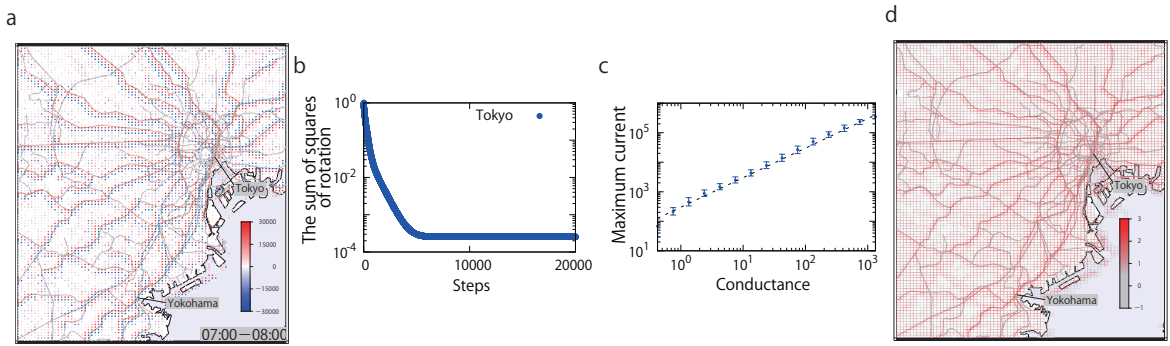}

\caption{\textbf{Optimization process and estimated conductance around Tokyo metropolitan area.} (a) Rotation patterns on the map in the morning. (b) Step-by-step decrease in the sum of squares of rotation calculated in the Tokyo metropolitan area. (c) Relationship between the maximum current at each cell for a day and the estimated conductance with a scaling exponent of 1.0.(d) Volume map of conductance in the Tokyo metropolitan area drawn on a logarithmic scale.}
\label{fig:fig2}
\end{figure*}

Fig. \ref{fig:fig2}a shows the spatial configuration of rotation for a morning rush hour at the initial iteration $\tau=0$ (i.e., the case of uniform resistance). We find that high rotation values of opposite signs are located along the railways, which are caused by large currents. By the above optimization process, the resistance near high current parts was reduced to decrease the appearance of rotation pairs with opposite signs. As shown in Fig. \ref{fig:fig2}b the sum of squares of rotation decreases quickly to less than 1 / 1000 after approximately 5000 iterations, thus the values of resistance converge. Fig. \ref{fig:fig2}c shows the relationship between the maximum current at each cell for a day and the estimated conductance. Conductance, $G=1/R$, is proportional to the maximum current at each cell, which can be thought to represent the capacity of railroads and highways. Fig. \ref{fig:fig2}d shows the spatial configuration of estimated conductance values for the same area as Fig. \ref{fig:fig2}a. We confirmed that the conductance values are high along the railways and highways.

\begin{figure*}

\centering

\includegraphics{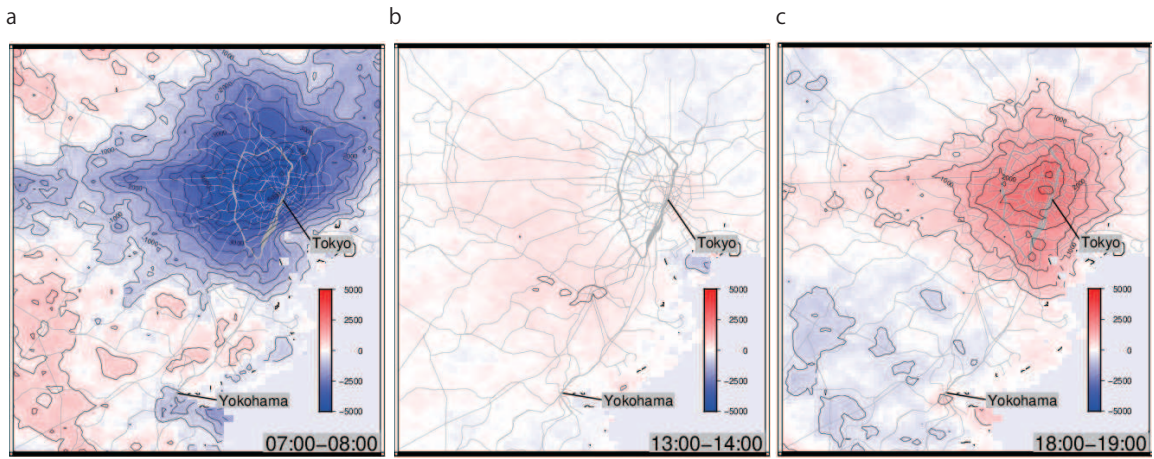}

\caption{\textbf{Temporal changes of electric potential (a) Morning, (b) Afternoon, (c) Evening.} Blue and red represent valley-like and mountain-like shapes, respectively.}
\label{fig:fig3}
\end{figure*}

%%\subsection{POTE}
As shown, a nearly rotation-free electric field $\bm{I}R$ is achieved by optimizing conductance. We calculate the electric potential by solving Poisson's equation. The discretized Poisson's equation of the electric network is given as: 
\begin{eqnarray}
[\nabla\cdot {\bm{I}R}]_{i,j,k}&=&Q_{i,j,k}  \\
\bm{I}_{(i,j)\rightarrow(i+1,j),k}R_{(i,j),(i+1,j)}&=&\phi_{i,j,k}-\phi_{i+1,j,k},\nonumber
\label{eq:Poisson}
\end{eqnarray}
where $Q_{i,j,k}$ denotes the sink or source of charges in cell $(i,j)$ at time interval $k$, and $\phi_{i,j,k}$ is the electric potential of the said cell. The value of the electric potential on the boundary of the model was set to 0 (see Supplementary for  boundary condition influence). For a given $Q_{i,j,k}$, the value of $\phi_{i,j,k}$ is uniquely determined by solving the linear equation (4). The morning and evening estimated potentials around the Tokyo metropolitan area are valley- and mountain-shaped, respectively, centered in central Tokyo (see Figs.\ref{fig:fig3}a and \ref{fig:fig3}c). These results are consistent with human flow patterns in commuting and returning rush characterized by directed drainage basins\cite{shida2020universal}. In contrast, the potential in the afternoon (see Fig. \ref{fig:fig3}b) is nearly flat, which is consistent with the result that daytime flow patterns are approximately random\cite{shida2020universal}.

\begin{figure*}

\centering
\includegraphics{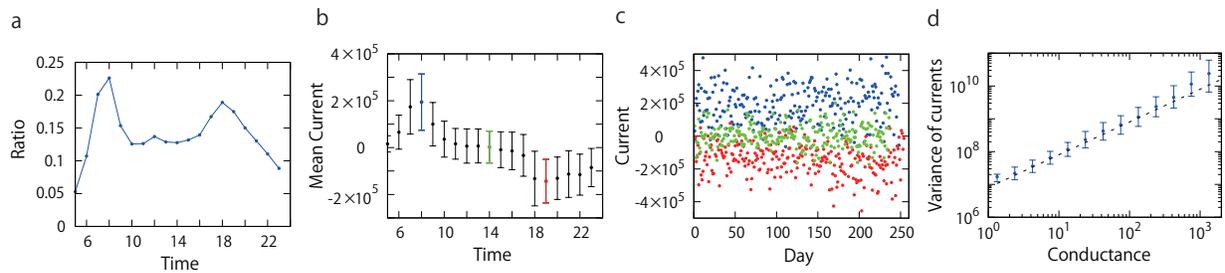}

\caption{\textbf{The relations between the variance of current and estimated conductance around Tokyo metropolitan area.} (a) Ratio of moving people in the Tokyo metropolitan area by time of day. (b) Example of average current change in a cell during the day. Error bars represent standard deviation. (c) Weekday currents in the morning (blue), afternoon (green), and evening (red) of the cells noted in Fig. \ref{fig:fig4}b are plotted daily. (d) Relationship between the variance of the current in each cell and the conductance with scaling exponent 1.0.}
\label{fig:fig4}
\end{figure*}

%%\subsection{Fluctuation of current}
Next, we focus on the number of moving people who contributed to the electric field. Fig. \ref{fig:fig4}a shows the time evolution of the moving people to userbase ratio. In the morning peak hour, about 23\% of people are moving while about 12\% are moving in the afternoon and about 18\% in the evening rush hour. Since the movement of people in the afternoon is approximately random\cite{shida2020universal}, we may assume that the 6-11\% difference is people's coherent motion that generates electric potential.

At the end of this study, we focus on fluctuations around the mean currents. Fig. \ref{fig:fig4}b represents the mean currents and the standard deviation of daily fluctuation of currents for each time interval observed at a resistor located between the residential area and the city center along a major railway. Fig. \ref{fig:fig4}c shows the daily values of currents during weekdays for the said resistor at three time intervals: morning rush hour, afternoon, and evening rush hour. We find large fluctuations in all cases, which is consistent with Fig. \ref{fig:fig4}b, because the standard deviations are nearly constant for all time intervals. Fig. \ref{fig:fig4}d is a log-log plot of the variance of current fluctuations conditioned by the conductance value $G$ for all resistors for all time intervals. We find a linear relation between the variance of currents and the conductance, which is a familiar relation known as the fluctuation-dissipation relation for resistors in thermal equilibrium with no mean current\cite{nyquist1928thermal}. However, our imaginary electric circuit model is far from thermal equilibrium. This relation may hold because there are about 10\% of people who move nearly randomly, as typically seen in the afternoon, which may mainly contribute to the fluctuation-dissipation relation. These results were also confirmed for two other urban areas, Osaka and Nagoya (see Supplementary).

%%\section{Conclusion}
In summary, we established a basic framework for mapping human flow observed by GPS data in urban areas to an imaginary electric circuit network reflecting the transportation infrastructure. This electrical circuit model can be applied to GPS data for any city in the world. Our model may provide the platform of numerical simulation of human flow of metropolitan area considering both synchronized potential flow and random fluctuations. Especially during the pandemic, our method may offer invaluable insight for policy-making in managing human flows.

\begin{acknowledgments}
We thank Agoop for providing GPS datasets. This work was supported by the Tokyo Tech World Research Hub Initiative (WRHI) Program of the Institute of Innovative Research, Tokyo Institute of Technology. This work was supported by a Grant-in-Aid for Scientific Research (B), Grant Number 18H01656. We would like to thank Editage (www.editage.com) for English language editing.

M.T. was the leader of this project, designed the whole research plan and directed writing of the manuscript. Y.S. analyzed the raw data, performed the numerical calculations, and wrote the manuscript. H.T. and J.O. developed methods of data analysis and revised the manuscript.
\end{acknowledgments}


\bibliography{main}% Produces the bibliography via BibTeX.

\end{document}